\documentclass[conference, letterpaper]{IEEEtran}
\IEEEoverridecommandlockouts    

\usepackage{multirow}
\usepackage{lipsum}
\usepackage{amsmath}
\usepackage{amssymb}
\usepackage[ruled,vlined]{algorithm2e}
\usepackage{graphicx}
\usepackage[pdftex,colorlinks=true,linkcolor=black,urlcolor=blue,citecolor=black]{hyperref}
\usepackage[section]{placeins}
\graphicspath{{./Figures/}}

\title{\LARGE \bf Open-Source METANET Calibration for Reproducible Freeway Traffic Macroscopic Simulation}

\author{
	\parbox{\textwidth}{%
		\centering
		Monica Chan$^{1}$, Shreyaa Raghavan$^{2}$, Cathy Wu$^{2}$%
	}%
	\thanks{$^{1}$Monica Chan (mochan@mit.edu) is with the Department of Electrical Engineering and Computer Science (EECS) at the Massachusetts Institute of Technology (MIT).  }
	\thanks{$^{2}$Shreyaa Raghavan (shreyaar@mit.edu) and  Cathy Wu (cathywu@mit.edu) are with the Institute for Data, Systems, and Society (IDSS) at MIT.}
    \thanks{Accepted at the IEEE Intelligent Transportation Systems Conference
(ITSC) 2026.}%
}

\begin{document}
	
	\maketitle
	\thispagestyle{empty}
	\pagestyle{empty}
	
	\begin{abstract}
		METANET is a widely used second-order macroscopic traffic flow model for freeway networks, supporting applications across traffic simulation, ramp metering, and variable speed limit control. The predictive accuracy of any traffic model, however, hinges on careful calibration to real-world conditions. Despite its widespread use, there have not been open-source tools for calibrating METANET's parameters. Without open-source calibration, results cannot be easily reproduced or extended to other networks. This work provides an open-source METANET calibration, simulation, and data visualization tool. The calibration is formulated as a nonlinear program (NLP) solved via the interior-point method (IPOPT), with joint ramp flow estimation. We validate our calibration on real-world freeway data from two widely used traffic monitoring systems:  Interstate-24 MObility Technology Interstate Observation Network (I-24 MOTION), one of the largest open-road trajectory instruments in the country, and loop detector data from the Caltrans Performance Measurement System (PeMS), which spans nearly 40,000 detectors across California freeways and serves as a standard benchmark in traffic research. Models calibrated using our method are able to reproduce these datasets' observed traffic patterns across diverse network geometries and traffic conditions including complex stop-and-go congestion waves. As large-scale traffic monitoring infrastructure continues to expand, open-source calibration tools are essential for translating growing volumes of sensor data into validated models that can support real-world traffic control. The complete code is publicly available at \href{https://github.com/woxsao/metanet-calibration}{this Github repository} to support reproducible research in freeway traffic modeling and control.

	\end{abstract}
	
	\section{Introduction}
	\label{sec:introduction}
	Freeway traffic modeling is a vital tool used for transportation research, helping researchers quantify how control methods such as variable speed limits (VSL) and ramp metering affect freeway congestion. Macroscopic models that utilize aggregate flows and speeds are useful because of their computational efficiency compared to microscopic trajectory-level simulators which are slower at scale \cite{Messmer_Papageorgiou_1990}. However, for these macroscopic simulators to provide meaningful insight, they require calibration so that their behavior reflects the real-world traffic conditions of the roadway being studied. 

    METANET is a discrete-time, second order macroscopic traffic flow model for freeway networks \cite{Messmer_Papageorgiou_1990}. METANET models freeway networks as directed graphs of links and nodes. It captures overall flows, speeds, and densities at points along the network. The dynamics equations capture relaxation, convection, and anticipation effects. In order to capture these dynamics, METANET requires a set of parameters for the underlying equations (relaxation time constant $\tau$, anticipation coefficient $\eta$, smoothing constant $\kappa$) and fundamental diagram (free flow velocity $v^*$, critical density $\rho^*$, model exponent term $a$). Determining these parameters is a key problem to solve in order to accurately model real road networks. 
    
    METANET has been used to evaluate techniques like model predictive control (MPC) for ramp metering \cite{Hegyi_2004}. More recent attempts that utilize METANET for ramp metering via reinforcement learning \cite{Airaldi_Schutter_Dabiri_2025} demonstrate METANET's ongoing relevance for transportation control research. Notably, neither study validates on real-world data, and \cite{Hegyi_2004} explicitly identifies calibration as a direction for future work. An open-source calibration tool would make it straightforward to ground such control studies in real freeway traffic data.

    Prior work on METANET calibration employs gradient based methods like RPROP \cite{Poole_Kotsialos_2016}, NLP gradient based optimization \cite{Frejo_Camacho_Horowitz_2012}, and cross entropy methods \cite{Ngoduy_Maher_2012}, but these methods do not have publicly available code, which makes replicating their results difficult. While open-source METANET implementations exist, such as sym-metanet \cite{Airaldi_2025}, they provide simulation only and lack calibration capabilities. Without shared calibration tools, each research team must independently implement their own calibration procedure, making it difficult to reproduce results, compare methods, or build on prior work.

    Moreover, most METANET formulations assume ramp flows are known inputs, yet ramp detector data is frequently unavailable or unreliable in practice \cite{Muralidharan_Horowitz_2009}, \cite{Li_Chen_Li_Gao_Li_Zhang_Dong_2025}. As a result many existing calibration works such as \cite{Poole_Kotsialos_2016}, \cite{Frejo_Camacho_Horowitz_2012}, \cite{Ngoduy_Maher_2012} which require ramp flow measurements as exogenous inputs are limited to corridors with comprehensive detector coverage. Our formulation overcomes this limitation by jointly estimating ramp flows and model parameters from mainline detector data alone.

    This work contributes 1) an open-source METANET simulation and calibration framework in Python, publicly available on \href{https://github.com/woxsao/metanet-calibration}{Github}, and 2) validation of the framework on a variety of real freeway networks, including trajectory level data from the I-24 MOTION system and loop detector data from the Caltrans Performance Measurement System (PeMS). This validation work demonstrates that our open-source framework can be used to calibrate to traffic patterns across different data sources, network geometries, and traffic conditions. We envision this tool will be useful to transportation researchers studying traffic control methods such as VSL or ramp metering by reducing the time and effort into developing a custom calibration and simulation program.

    The remainder of this paper is organized as follows. \autoref{sec:metanetformulation} presents the METANET model formulation. \autoref{sec:calibrationmethodology} describes the calibration methodology. \autoref{sec:experimentaldesign} describes the validation experimental design including data used. \autoref{sec:results} discusses results on those experiments, followed by conclusions in \autoref{sec:conclusion}. 

	\section{METANET model Formulation}
	\label{sec:metanetformulation}
    
	The modeled freeway corridor is discretized into N segments of equal length $L$, as shown in \autoref{fig:network-example}. Each segment $x$ is assigned a lane count $\lambda_x$, which may vary across segments to represent lane drops and additions. 
    
    The traffic state at each segment $x$ and time step $t$ is described by total density $\rho_{t,x}$ (veh/km), total flow $q_{t,x}$ (veh/hr), and mean speeds $v_{t,x}$ (km/hr). Per lane densities can be found by $\rho_{t,x}/\lambda_x$. Effects of on ramp and off-ramps are represented by a per-segment inflow value $r_{t,x}$ and outflow splitting ratio $\beta_{t,x}$, which are nonzero only at segments where physical ramps are present. In the example of \autoref{fig:network-example}, $r_{t,x} = 0 \quad \forall \, t, x$, and $\beta_{t,4} \geq 0 \quad \forall \, t$. Boundary conditions are required at the upstream and downstream ends of the corridor: upstream demand entering the first segment and downstream density exiting the last segment. The dynamics at each segment can be described by the following equations:

    \begin{figure}
        \centering
        \includegraphics[width=0.9\linewidth]{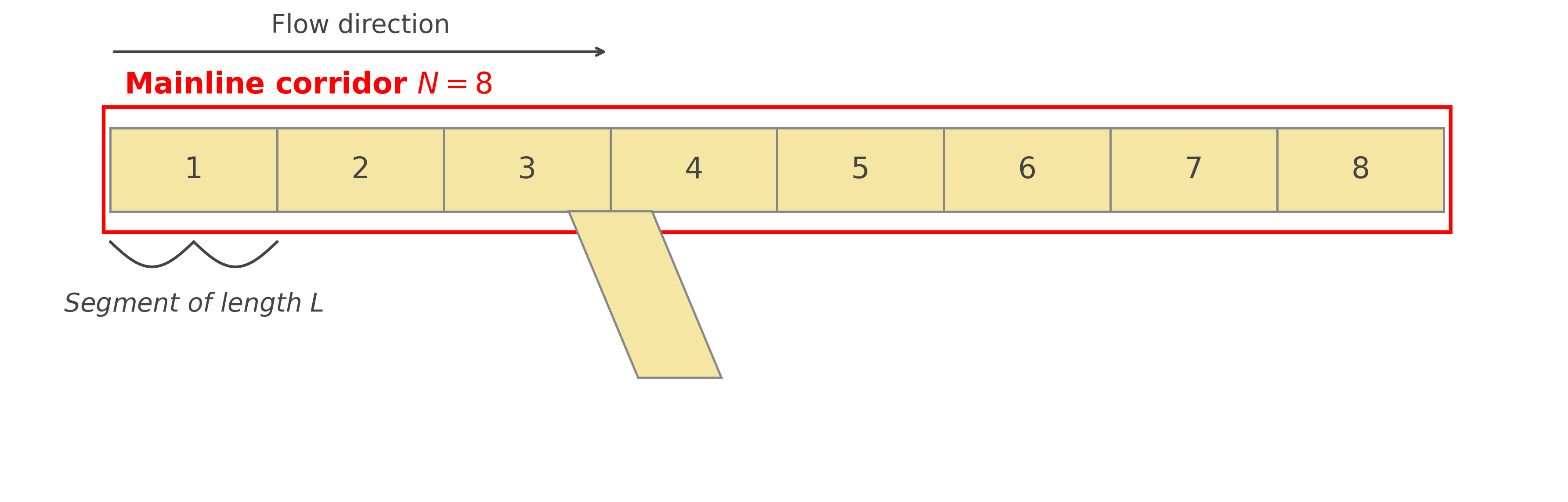}
        \caption{Example of network representation. The mainline corridor is split into $N$ segments of length $L$}
        \label{fig:network-example}
    \end{figure}
	
	\begin{align}
    \rho_{t+1, x} &= \rho_{t, x} + \frac{\delta}{L} \left(q_{t,x-1} - \frac{q_{t,x}}{1 - \beta_{t,x}} + r_{t,x} \right) \label{density} \\
    q_{t,x} &= \rho_{t,x} v_{t,x} \label{flow} \\
    v_{t+1,x} &= v_{t,x} + \frac{\delta}{\tau_x} (V[\rho_{t,x}] - v_{t,x})   \notag \\
    &   + \frac{\delta v_{t, x} }{L} \left( v_{t,x-1}- v_{t, x}\right) 
    - \frac{\eta_x \delta}{\tau_x L} \frac{\frac{\rho_{t,x+1}}{\lambda_{x+1}} - \frac{\rho_{t,x}}{\lambda_x}}{\frac{\rho_{t,x}}{\lambda_x} + \kappa_x}
    \label{velocity} \\
    V[\rho_{t,x}] &= v^*_{x} \exp \left[-\frac{1}{\alpha}\left(\frac{\rho_{t,x}}{\rho^*_{x} \lambda_x}\right)^{\alpha}\right]
    \end{align}

    where $\delta$ is the simulation timestep. The model parameters to be calibrated are the relaxation time constant $\tau$, anticipation coefficient $\eta$, smoothing constant $\kappa$, free flow speed $v^*$, critical density $\rho^*$, and model exponent $\alpha$.

    Most METANET formulations, including the original by Messmer and Papageorgiou, represent the freeway as a directed graph of links connected by nodes, where on-ramps and off-ramps are modeled as separate links with known inflow and outflow boundary conditions \cite{Messmer_Papageorgiou_1990}. The assumption for such a formulation is that the inflows of on-ramps and off-ramps are provided. However, prior work has noted that ramp inflow and outflow data is often missing or unreliable \cite{Muralidharan_Horowitz_2009}, \cite{Li_Chen_Li_Gao_Li_Zhang_Dong_2025} which aligns with our experience. To address this limitation, our formulation instead models the freeway as a sequence of $N$ segments, each with its own traffic state and parameter set. On-ramp flows $r_{t,x}$ and off-ramp split ratios $\beta_{t,x}$ enter directly as source and sink terms in each segment's conservation equation. These flows and split ratios can be treated as either decision variables in the calibration or user-provided inputs. This formulation resolves the missing ramp data issue. 
    
	\section{Calibration Methodology}
    \label{sec:calibrationmethodology}

    The proposed approach formulates METANET dynamics as an NLP using Pyomo \cite{Bynum_Hackebeil_Hart_Laird_Nicholson_Siirola_Watson_Woodruff_2021}, solved with the interior point solver IPOPT \cite{Wachter_Biegler_2006}. The decision variables are the model parameters themselves $(\tau_x, \eta_x, \kappa_x, v^*_x, \rho^*_x, \alpha_x, r_{t,x}, \beta_{t,x})$. The METANET dynamics \autoref{density}, \autoref{flow}, and \autoref{velocity} are enforced as equality constraints, ensuring the simulated traffic states satisfy the model at every segment and time step. This method is similar to prior work utilizing MATLAB optimizing functions \cite{Frejo_Camacho_Horowitz_2012}.
    
    Each segment is assigned an independent set of parameters, which allows the calibration to capture spatial heterogeneity along the corridor.  For long corridors, the network can be calibrated in blocks of consecutive segments, with observed data from adjacent segments providing boundary conditions.
    
    \subsection{Objective Function}
    \label{subsec:objectivefunc}

    While the objective function can be adapted to specific applications, we adopt a weighted, normalized sum of squared errors between simulated and observed traffic states. 
    \begin{align}
        \min \sum_{t} \sum_{x} \Bigg[ w_v \left(\frac{v_{t,x} - \bar{v}_{t,x}}{\bar{v}_{\max}}\right)^2 &+ w_\rho \left(\frac{\rho_{t,x} - \bar{\rho}_{t,x}}{\bar{\rho}_{\max}}\right)^2 \notag \\
        &+ w_q \left(\frac{q_{t,x} - \bar{q}_{t,x}}{\bar{q}_{\max}}\right)^2 \Bigg] \label{eq:objective}
    \end{align}
    where $\bar{v}, \bar{\rho}, \bar{q}$ represent the ground truth speeds, densities and flows respectively. Each term is  normalized by its observed maximum. For experiments in this paper, $w_v=20, w_\rho = 1, w_q = 1$.

    \subsection{Data Preparation}
    \label{subsec:dataprep}

    Ground truth data should be prepared at the desired segment resolution. That is, if calibrating to an hour's worth of data of a 4 km network with $\delta=10/3600$ hours and $L=0.4$km, the calibration technique expects aggregate flows, densities, and velocities arrays of shape (timesteps, segments) = (360, 10). If separate boundary conditions of upstream flow and downstream density are not provided, the calibration technique will use the first and last segment for the upstream inflow and downstream density, respectively, and will only calibrate to the interior segments.

    Bounds of the decision variables can be customized on input. Default values are provided in \autoref{tab:defaultbounds}. They are loosely based on commonly used ranges in METANET literature \cite{Poole_Kotsialos_2016}, \cite{Frejo_Camacho_Horowitz_2012}. These bounds worked well across scenarios tested in this paper, but users might need to adjust to suit their corridor. For example, $v^*$ bounds may need to increase for higher speed freeways, and $r$ bounds should reflect expected ramp volumes. A one-hot encoding of on ramp and off-ramp locations is expected. Ramp inflow and turning ration estimates $r_{t,x}$ and $ \beta_{t,x}$ are constrained to only be nonzero at segments where there are ramps. Specification of number of lanes per segment is also expected. 
	
    \begin{table}[h]
    \centering
    \caption{Default Variable Bounds}
    \label{tab:defaultbounds}
    \begin{tabular}{lcc}
    \hline
    \textbf{Parameter} & \textbf{Lower} & \textbf{Upper} \\
    \hline
    $\tau$ (hours) & $15/3600$ & $60/3600$ \\
    $\eta$ & 15 & 60 \\
    $\kappa$ & 5 & 60 \\
    $v^*$ (km/h) & 110 & 150 \\
    $\rho^*$ (veh/km/lane) & 15 & 100 \\
    $\alpha$ & 0.5 & 5 \\
    $\beta$ & 0 & 0.9 \\
    $r$ (veh/h) & 0 & 2000 \\
    \hline
    \end{tabular}
    \end{table}

    \subsection{Solver Configuration}
    The user can specify parameters such as max iterations and convergence tolerances, but the default is configured with a maximum of 20,000 iterations and a constraint violation tolerance of $10^{-13}$ to ensure convergence. Since the calibration program is written using Python's Pyomo library, configurations such as the solver, tolerances, and maximum iterations can be customized to match user requirements. Default variable initialization is taken as the midpoint of the bounds, but can be configured.
    
    We note that IPOPT solutions may vary across machines due to differences in floating-point arithmetic, linear algebra libraries (e.g., MUMPS vs. MA57), and compiler optimizations. Because the calibration problem is non-convex, these small numerical differences can lead the solver down different paths, potentially converging to different local minima. Users should therefore expect minor differences in calibrated parameters when running on different computers.
    
	\section{Experimental Design}
    \label{sec:experimentaldesign}

\begin{figure}
        \centering
        \includegraphics[width=0.9\linewidth]{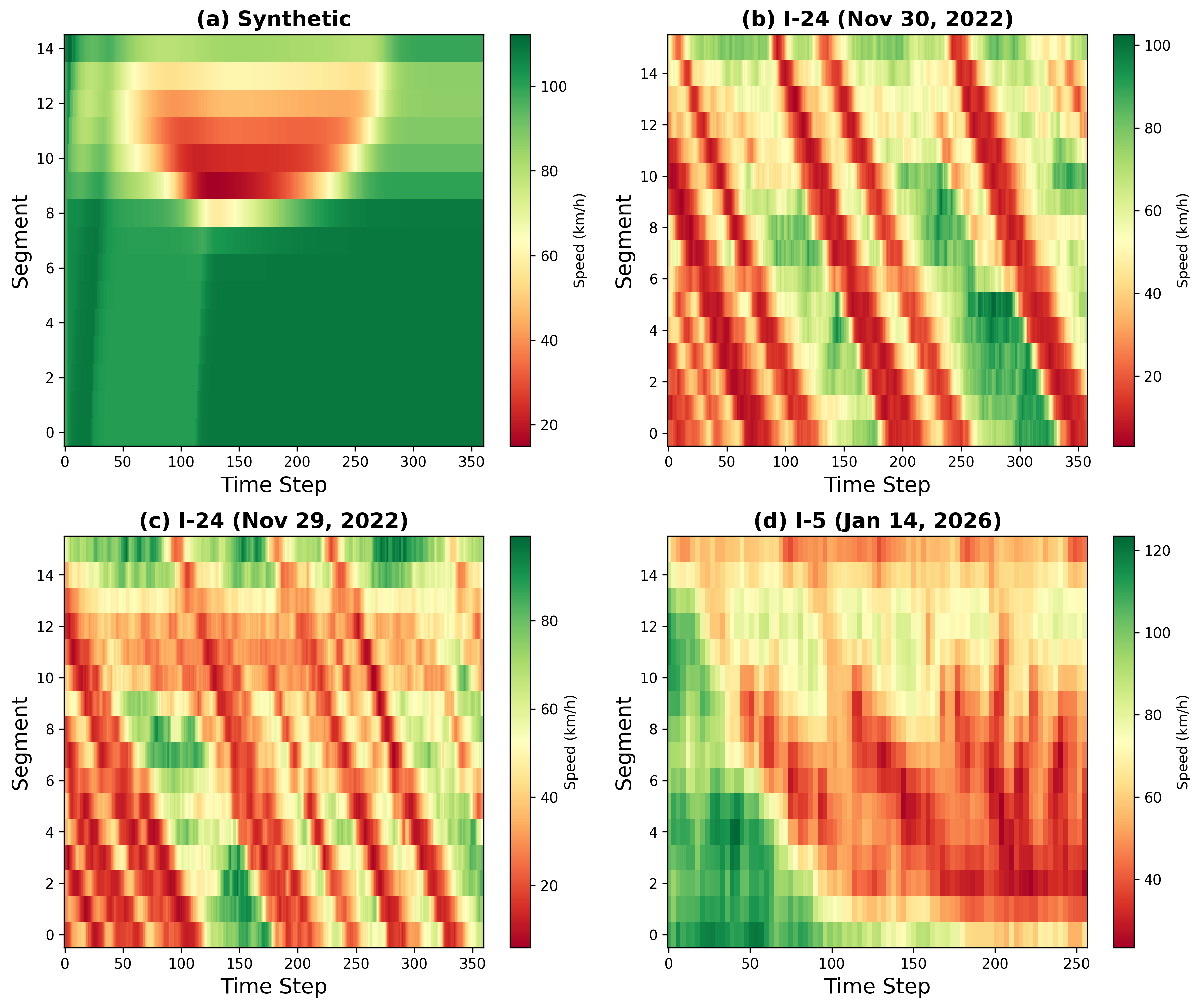}
        \caption{Speed time-space diagrams of the validation set of networks. a. is the synthetic bottleneck, b. is I-24 MOTION on 11-30-2022 from 8-9AM, c. is I-24 MOTION on 11-29-2022 from 8-9AM, and d. is I-5 from PeMS on 1-14-2026 from 7-8AM.}
        \label{fig:eval_set}
    \end{figure}

    This calibration technique was tested on four scenarios: a synthetic bottleneck scenario, two separate hours of  Interstate-24 MObility Technology Interstate Observation Network (I-24 MOTION) data, and 42 minutes of Caltrans Performance Measurement System (PeMS) Interstate 5 (I-5) data. The speed profile traffic pattern can be seen in \autoref{fig:eval_set}. The goal of the validation process is to report on the mean average percent error (MAPE) between the simulated traffic state and the ground truth traffic state.  

    Lane counts, as shown in \autoref{fig:i24-schematic} and \autoref{fig:i5-schematic} are not necessarily integer values. When a lane drop or addition occurs within a segment, the lane count is computed as a weighted average over the segment length. For example, if a segment is 400m long and the first 300m have 4 lanes and the remaining 100m has 5 lanes, the segment is assigned 4.25 lanes.
    
    \subsection{Synthetic Scenario}
\begin{figure}
        \centering
        \includegraphics[width=0.9\linewidth]{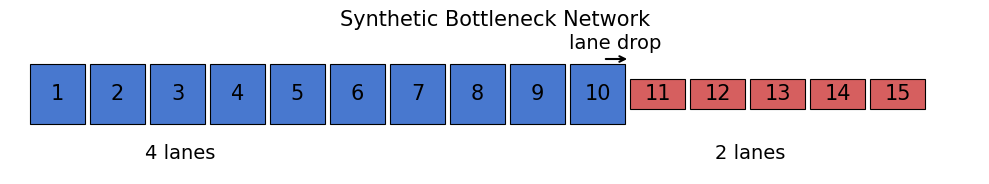}
        \caption{Synthetic bottleneck network. Segments 1–10 have 4 lanes; a lane drop to 2 lanes occurs between segments 10 and 11. Each segment is 0.4km.}
        \label{fig:synthetic-network}
    \end{figure}

    \autoref{fig:synthetic-network} shows a diagram of the synthetic network. Between segments 10 and 11 there is a lane drop from four to two lanes to create congestion. The ground truth parameters that generate the data are shown in \autoref{tab:synthetic_params}. Each segment has the same parameters. 
    
    \begin{table}[h]
    \centering
    \caption{Ground Truth Parameters for Synthetic Scenario}
    \label{tab:synthetic_params}
    \begin{tabular}{lcc}
    \hline
    \textbf{Parameter} & \textbf{Symbol} & \textbf{Value} \\
    \hline
    Relaxation time & $\tau$ & $18/3600$ hr \\
    Anticipation constant & $\eta$ & 30 \\
    Model constant & $\kappa$ & 40 \\
    Free-flow speed & $v_{\text{free}}$ & 120 km/hr \\
    Critical density & $\rho_{\text{crit}}$ & 37.45 veh/km/lane \\
    Exponent & $\alpha$ & 1.4 \\
    On-ramp flow & $r$ & 0 veh/hr \\
    Off-ramp split ratio & $\beta$ & 0 \\
    \hline
    \end{tabular}
    \end{table}

    \subsection{I-24 Motion Data}
\begin{figure}
        \centering
        \includegraphics[width=1.0\linewidth]{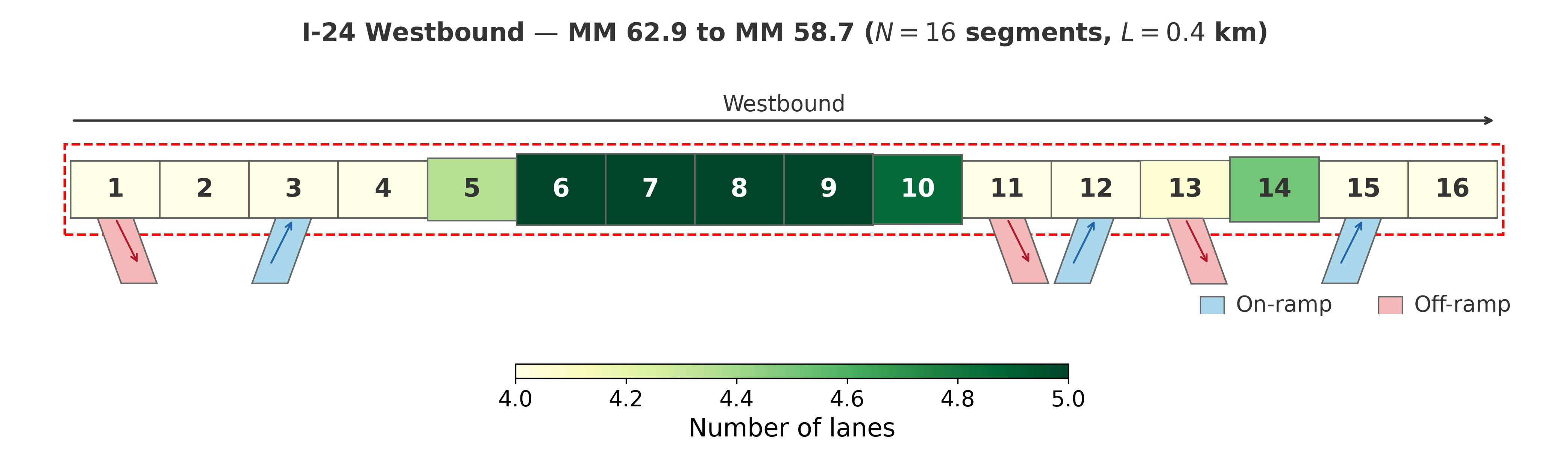}
        \caption{Macroscopic representation of the I-24 main corridor used for calibration. Each segment is 0.4km. Note that the actual corridor is not geometrically straight but is shown straight for modeling purposes.}
        \label{fig:i24-schematic}
    \end{figure}

    We calibrated to two separate hours on two separate days of the I-24 MOTION dataset \cite{Gloudemans_Wang_Ji_Zachar_Barbour_Work_2023}. I-24 MOTION consists of trajectory level microscopic data collected by camera poles along a 4.2 mile stretch of Interstate 24 near Nashville, TN. We aggregate this data into macroscopic flows, densities, and speeds. This corridor exhibits recurring stop-and-go waves, providing a challenging test of whether the calibrated model can reproduce complex congestion dynamics. We calibrated to the westbound data because its network geometry creates the waves shown in \autoref{fig:eval_set}b-c. The macroscopic schematic of the network is shown in \autoref{fig:i24-schematic}. Both hours of data capture westbound traffic on I-24 from 8 to 9 am local time.

    Since I-24 MOTION data is at the trajectory level, it is aggregated to macroscopic level using Edie's definitions \cite{Edie_1963}. The corridor is discretized into space-time cells of length $L = 0.4$km and $\delta = 10/3600$hr. For each cell, flow and density are computed as 
    \begin{equation}
        q = \frac{\sum_i d_i}{L\cdot \delta} \qquad \rho = \frac{\sum_it_i}{L\cdot \delta}
    \end{equation}

    where $d_i$ is the distance traveled by vehicle $i$ within the cell and $t_i$ is the time spent. Speed is then $v = q/\rho$.
    \subsection{I-5 PeMS Data}
\begin{figure}
        \centering
        \includegraphics[width=1.0\linewidth]{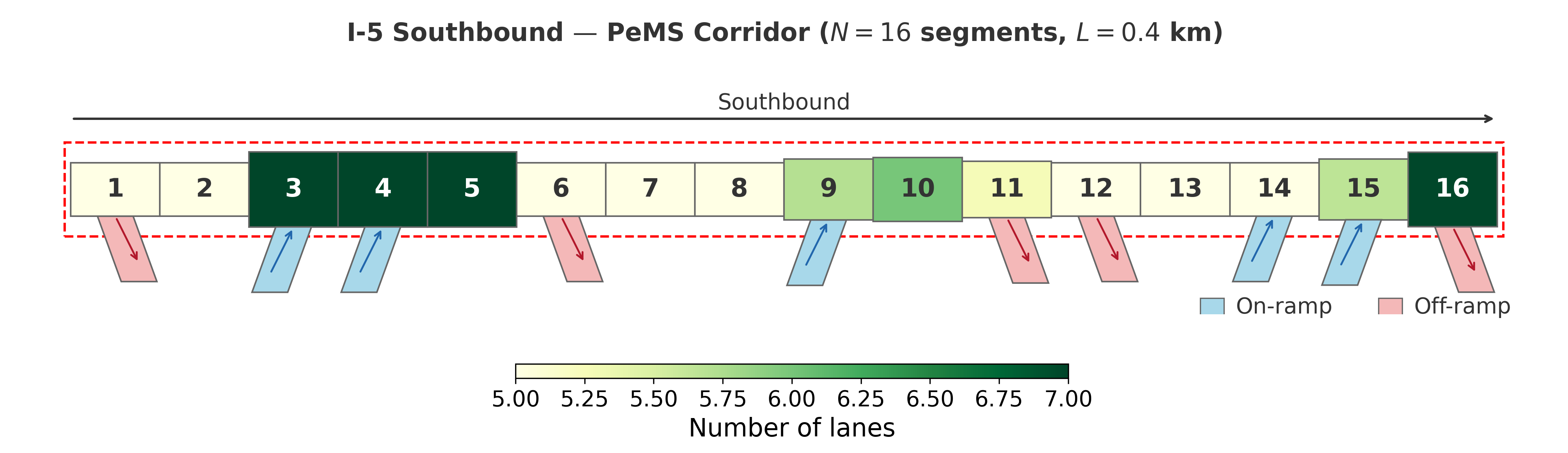}
        \caption{Macroscopic representation of the I-5 main corridor used for calibration. Each segment is 0.4km. As is with \autoref{fig:i24-schematic}, the corridor is not as straight as pictured.}
        \label{fig:i5-schematic}
    \end{figure}

\begin{figure}
        \centering
        \includegraphics[width=0.9\linewidth]{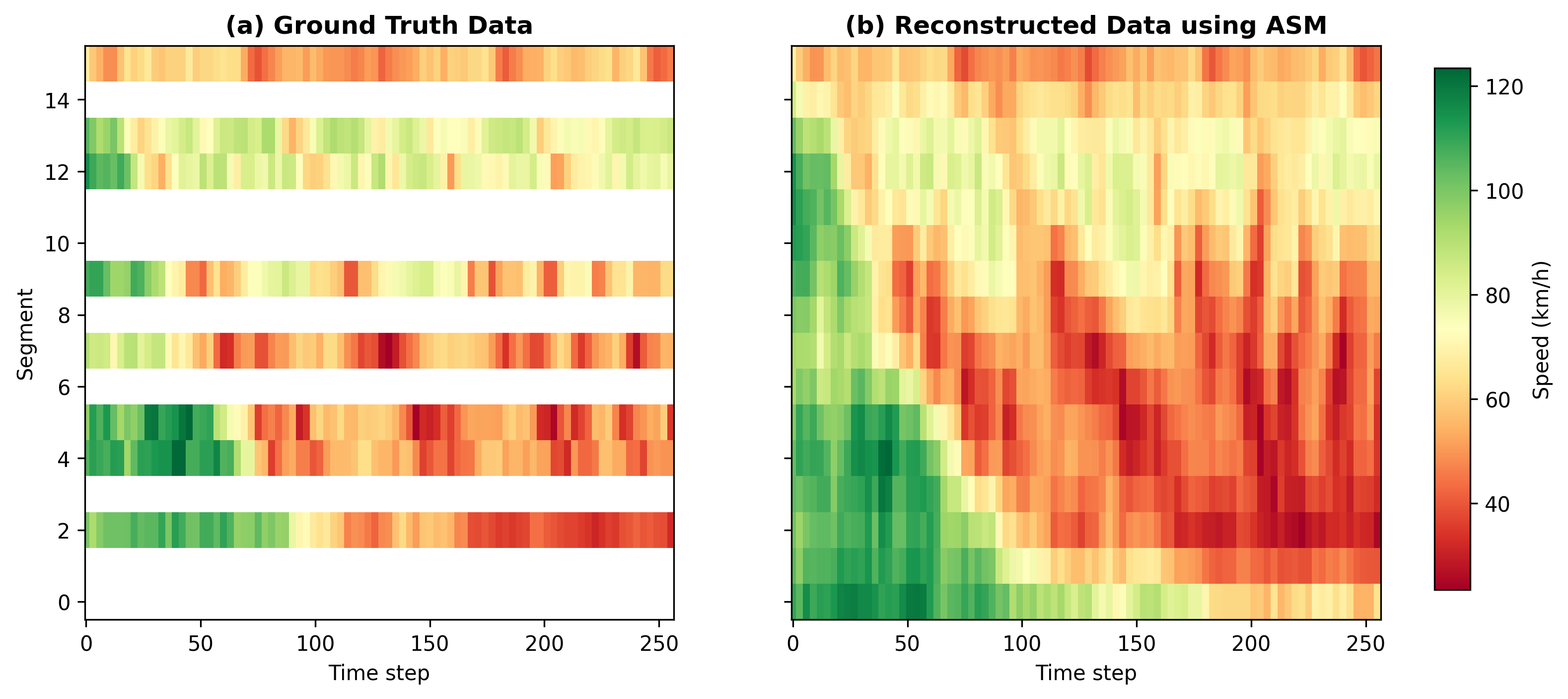}
        \caption{a) Original PeMS loop detector speed data and b) reconstructed missing segment data using ASM.}
        \label{fig:asm-reconstruction}
    \end{figure}

    PeMS is a widely used source of data, spanning multiple freeways in California \cite{Chen_2003}. PeMS provides 5 minute aggregated flow, speed, and occupancy data from loop detectors. We use flow and speed measurements and derive density as $\rho = q/v$. We include a PeMS corridor to demonstrate that the framework can calibrate directly from standard loop detector data without requiring trajectory level measurements. The segment of PeMS used can be seen in \autoref{fig:i5-schematic}, and it goes from mile marker 110A to 114B between Buena Park and Southeast Anaheim. We calibrate to 42 minutes of this data with $L=0.4$km, and $\delta =10/3600$ hours. Since PeMS data is from loop detectors, there are segments of our modeled network that do not have ground truth data. To provide a complete ground truth for evaluation, we reconstruct the missing data using the adaptive smoothing method (ASM) \cite{Ji_Gloudemans_Zachar_Nice_Barbour_Work_2026}. The original data vs. reconstructed data is shown in \autoref{fig:asm-reconstruction}.

	\section{Results}
    \label{sec:results}

    \begin{figure*}
        \centering
        \includegraphics[width=0.75\textwidth]{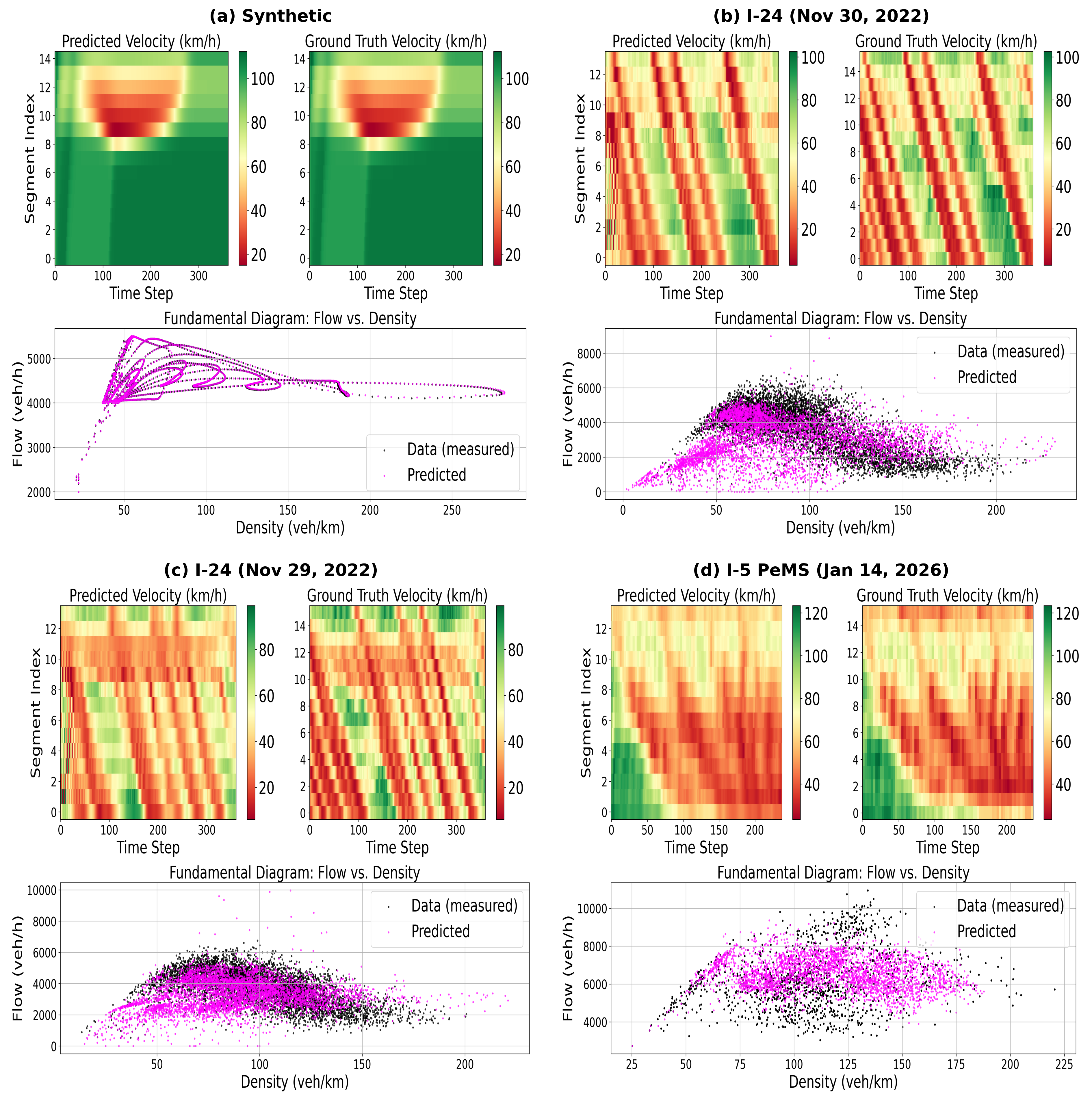}
        \caption{Results compared to ground truth data for a) synthetic scenario, b) I-24 on 11-30-2022, c) I-24 on 11-29-2022, and d) I-5. Each subplot compares predicted traffic state to the ground truth. The bottom scatter plot in each subplot shows the predicted (blue) and ground truth (gray) flows vs. densities. Each dot represents one time-space segment.}
        \label{fig:combined-results}
    \end{figure*}

\begin{table}[h]
    \centering
    \caption{Ablation Study: MAPE (\%) by Scenario and Model Configuration}
    \label{tab:results}
    \begin{tabular}{llcccc}
    \hline
    \textbf{Scenario} & \textbf{Config.} & $\rho$ & $q$ & $v$ & \textbf{Mean} \\
    \hline
    \multirow{2}{*}{Synthetic}     & SV (Ours) & 0.15 & 0.07 & 0.13 & 0.12 \\
                                   & SI        & \textbf{0.12} & \textbf{0.04} & \textbf{0.11} & \textbf{0.09} \\
    \hline
    \multirow{4}{*}{I-24 (Nov 29)} & SV-TV (Ours) & \textbf{23.71} & \textbf{27.63} & \textbf{26.88} & \textbf{26.07} \\
                                   & SI-TV        & 24.86 & 30.38 & 44.38 & 33.21 \\
                                   & SI-TI        & 28.73 & 31.52 & 56.44 & 38.90 \\
                                   & SV-TI        & 33.40 & 37.44 & 49.17 & 40.00 \\
    \hline
    \multirow{4}{*}{I-24 (Nov 30)} & SV-TV (Ours) & 28.64 & \textbf{34.55} & \textbf{28.30} & \textbf{30.50} \\
                                   & SI-TV        & \textbf{26.16} & 37.74 & 58.52 & 40.81 \\
                                   & SV-TI        & 35.51 & 54.22 & 63.92 & 51.22 \\
                                   & SI-TI        & 36.01 & 47.49 & 92.00 & 58.50 \\
    \hline
    \multirow{4}{*}{I-5 PeMS}      & SV-TV (Ours) & 25.46 & 27.70 & \textbf{5.44} & \textbf{19.53} \\
                                   & SV-TI        & \textbf{23.45} & \textbf{25.55} & 13.77 & 20.92 \\
                                   & SI-TV        & 28.19 & 30.44 & 7.05 & 21.89 \\
                                   & SI-TI        & 32.78 & 33.63 & 15.16 & 27.19 \\
    \hline
    \multicolumn{6}{l}{\footnotesize SI: seg.-invariant params, SV: seg.-varying params,} \\
    \multicolumn{6}{l}{\footnotesize TV: time-varying ramp, TI: time-invariant ramp.} \\
    \multicolumn{6}{l}{\footnotesize Note: The synthetic scenario has no ramps, so ramping ablation} \\
    \multicolumn{6}{l}{\footnotesize is not performed for the synthetic scenario.} \\
    \end{tabular}
\end{table}
    The resulting MAPEs for each scenario are listed in \autoref{tab:results}. \autoref{fig:combined-results} shows the predicted traffic states compared to the ground truth data. Each subplot shows the predicted speeds compared to the ground truth. In addition, each subplot includes a fundamental diagram scatter plot showing the predicted vs. ground truth flows. In all the cases we can see that our calibration method has recovered the underlying patterns including stop-and-go waves. 

    \autoref{tab:results} also reports an ablation over two design axes: segment-invariant
(SI) vs.\ segment-varying (SV) parameters, and time-invariant (TI) vs.\ time-varying
(TV) ramp flows. Time-varying ramps prove to be the more impactful choice: removing
them (SI-TI) causes the largest degradation in all scenarios, with velocity MAPE on
I-24 (Nov 30) increasing from 28.30\% to 92.00\%. Segment-varying parameters
provide additional benefit on I-24, where road heterogeneity is more pronounced,
but are less decisive on I-5 where geometry is more uniform. With this ablation study, the Total column shows that our proposed method, segment-varying parameters plus time-varying ramping gives us the lowest average MAPE across all 3 variables for all real data scenarios.

\begin{figure}
        \centering
        \includegraphics[width=1.0\linewidth]{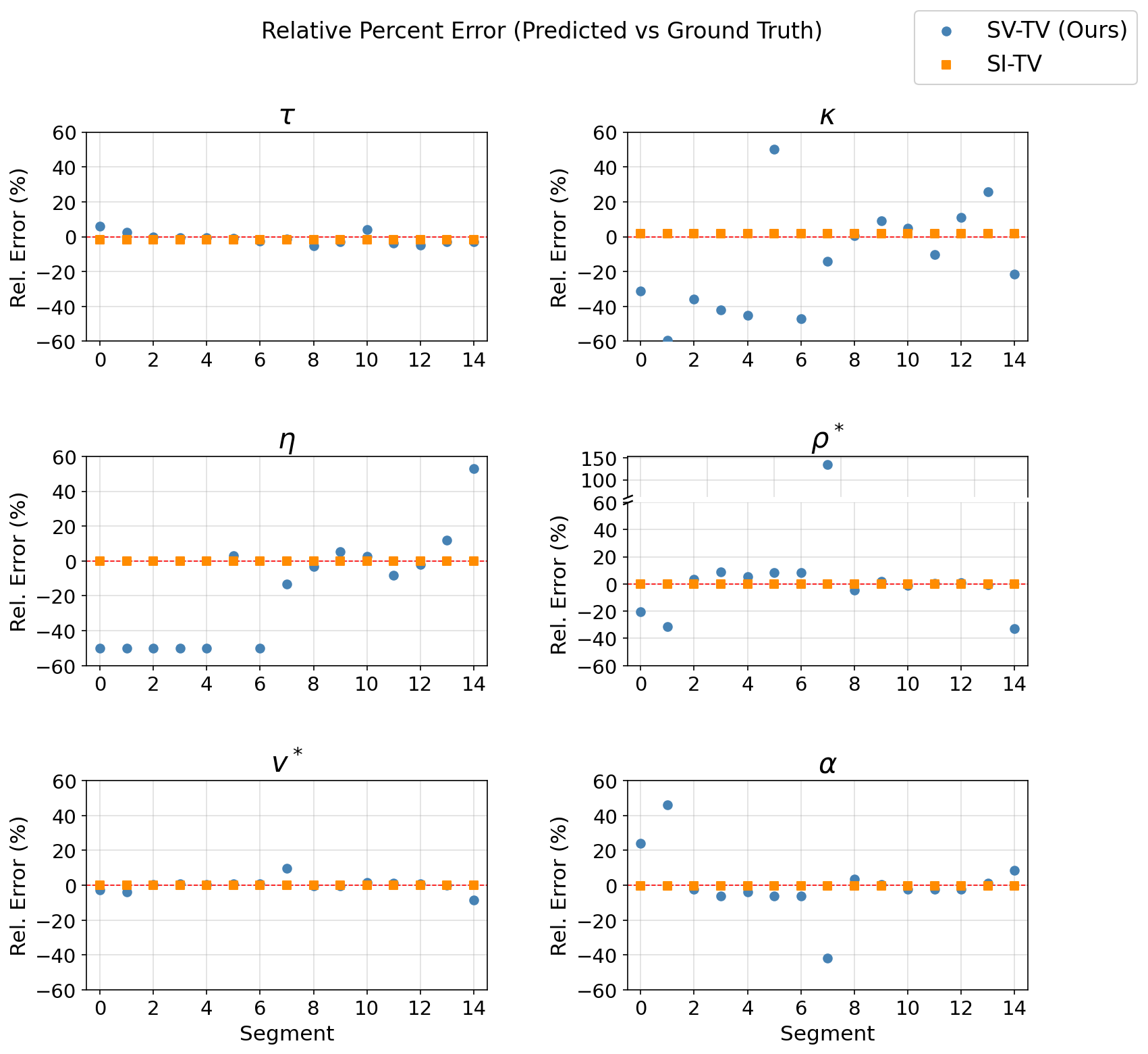}
        \caption{\textbf{Parameter Recovery: Synthetic Scenario.} Relative percent error between estimated and ground-truth METANET parameters. The synthetic scenario contains no on- or off-ramps, producing near-spatially-homogeneous dynamics; under these conditions SI-TV achieves lower MAPE than SV-TV (\autoref{tab:results}). All real-world scenarios include ramps, inducing spatial variation in traffic state that SI-TV's uniform parameterization cannot capture; SV-TV achieves lower MAPE in all real scenarios (\autoref{tab:results}).}
        \label{fig:parameter-differences}
    \end{figure}

    For the synthetic scenario, we were able to almost perfectly replicate the traffic pattern, but with somewhat different parameters, as seen in \autoref{fig:parameter-differences} for both the segment-varying and segment-invariant parameters. The model using segment-invariant parameters was much closer to the ground truth parameters. However, unlike the real-world networks used, the synthetic scenario is ramp-free and generated by underlying parameters that are the same across all segments, which yields dynamics that are spatially homogeneous. Under these conditions, segment-varying parameters introduces a needlessly complex optimization landscape. However, for all real data scenarios evaluated in this study, segment-varying parameters was necessary to capture the varying traffic states across the corridors introduced by ramps, road grade, etc. This suggests that when calibrating METANET, it is necessary to consider network geometry when deciding whether to vary parameters across segments. 

    For I-24 results, both scenarios' stop-and-go waves were captured by the calibrated parameters. The higher MAPE on this corridor reflects the difficulty of reproducing rapid oscillations between the free flow and congested states with time-invariant parameters. Between the two days of I-24 data, the MAPE is lower for Nov 29  than Nov 30, which we attribute to the more severe and irregular congestion observed on Nov 30. As can be seen in \autoref{fig:eval_set}, Nov 30 has noticeably more intense and sharper congestion. 

    The MAPE for velocity on the PeMS network is significantly lower than the flow or density MAPE, but this can be attributed to the weights for our objective function, since we weighted $w_v = 20, w_\rho =1, w_q = 1$. This high weight on the velocity term is primarily due to the fact that the speed update, \autoref{velocity}, contains most of the calibrated parameters ($v^*, \rho^*,\alpha, \tau, \eta, \kappa$), while density and flow are governed primarily by conservation. Weighting speed more heavily provides stronger gradient signals for the parameters during calibration.

	\section{Conclusion}
	\label{sec:conclusion}
    Macroscopic traffic models like METANET are widely used in traffic control research, yet calibrating them to real-world data --- which is crucial for the performance of these models --- remains a bottleneck in traffic analysis. Prior calibration methods lack publicly available code to reproduce results, and most require ramp detector data that is frequently unavailable. This work addresses both limitations with an open-source METANET calibration and simulation framework that formulates the calibration problem as an NLP solved with IPOPT. Unlike prior approaches, it jointly estimates model parameters and ramp flows from mainline data, removing the need for ramp detector coverage. The calibration and simulation are \href{https://anonymous.4open.science/r/metanet-calibration-733E/README.md}{publicly available on Github}, lowering the barrier for researchers to calibrate METANET to their own data.

    We validated our method on both synthetic and real data including I-24 MOTION and I-5 data from PeMS. Results from these experiments show that our calibration method is able to replicate real traffic dynamics including stop-and-go waves, and ablation studies confirm the performance benefits of jointly estimating ramp inflows and allowing parameters to vary across segments. While these results are promising, several open questions remain. First, a sensitivity analysis on the IPOPT initialization would clarify whether different starting points converge to the same solution, and whether the problem suffers from parameter non-identifiability. Second, examining sensitivity to the parameter bounds would reveal how much these constraints shape the parameters found by IPOPT. Finally, a more systematic approach to selecting objective function weights, which currently requires manual tuning for each network, would improve the generalization of this method to various highway systems.
	
	
    \FloatBarrier
	\bibliographystyle{IEEEtran}
	\bibliography{cite} 
	
\end{document}